\documentclass[conference]{IEEEtran}
\IEEEoverridecommandlockouts
\usepackage{amsmath,amssymb,amsfonts,amsthm}
\usepackage[ruled]{algorithm2e}
\usepackage{graphicx}
\usepackage{textcomp}
\usepackage{xcolor}
\usepackage[a4paper, total={184mm,239mm}]{geometry}

\usepackage{balance}
\usepackage{comment}
\usepackage{subfiles}
\usepackage{url}
\usepackage{enumitem}
\usepackage[braket, qm]{qcircuit}
\usepackage{subcaption}
\usepackage{pgfplots}
\pgfplotsset{compat=1.18}
\usepackage{color,soul}
\usepackage{csquotes}
\usepackage{todonotes}
\usepackage{booktabs}
\usepackage{algpseudocode}
\usepackage{multirow}
\usepackage{comment}
\usepackage{cite}
\usepackage{rotating}
\usepackage{afterpage}
\usepackage[export]{adjustbox}

\newcommand{\dirac}[1]{\ensuremath{\left|#1\right\rangle}} 

\def\BibTeX{{\rm B\kern-.05em{\sc i\kern-.025em b}\kern-.08em
    T\kern-.1667em\lower.7ex\hbox{E}\kern-.125emX}}
\begin{document}
\title{qSAT: Design of an Efficient Quantum Satisfiability Solver for Hardware Equivalence Checking}

\author{
    \IEEEauthorblockN{Abhoy Kole\IEEEauthorrefmark{1}  Mohammed E. Djeridane\IEEEauthorrefmark{2}\IEEEauthorrefmark{3} 
    Lennart Weingarten\IEEEauthorrefmark{3}  
    Kamalika Datta\IEEEauthorrefmark{1}\IEEEauthorrefmark{3} Rolf Drechsler\IEEEauthorrefmark{1}\IEEEauthorrefmark{3}}

    \IEEEauthorblockA{
    \IEEEauthorrefmark{1}Cyber-Physical Systems, DFKI GmbH, Bremen, Germany  \\
    \IEEEauthorrefmark{2}Siemens Electronic Design Automation GmbH, Hamburg, Germany  \\    
    \IEEEauthorrefmark{3}Institute of Computer Science, University of Bremen, Bremen, Germany  \\ 
    abhoy.kole@dfki.de,  \{djeridan,len\_wei,kdatta,drechsler\}@uni-bremen.de}
 }
\maketitle

\begin{abstract}

The use of Boolean Satisfiability (SAT) solver for hardware verification incurs exponential run-time in several 
instances. In this work we have proposed an efficient quantum SAT (qSAT) solver for equivalence checking of 
Boolean circuits employing Grover’s algorithm. The Exclusive-Sum-of-Product based generation of the Conjunctive 
Normal Form equivalent clauses demand less qubits and minimizes the gates and depth of quantum circuit 
interpretation. The consideration of reference circuits for verification affecting Grover’s iterations and 
quantum resources are also presented as a case study. Experimental results are presented assessing the benefits 
of the proposed verification approach using open-source Qiskit platform and IBM quantum computer.

\end{abstract}


\section{Introduction}

Real-world applications like hardware verification often pose a significant challenge for finding solutions in polynomial time. The verification problem is often formulated as Boolean Satisfiability (SAT) problem in the form of a \emph{miter} circuit for SAT solving using state of the art solvers like Z3 \emph{Satisfiability Modulo Theory} (SMT) solver~\cite{de2008z3}. The SAT problem is the first discovered NP-Complete problem\cite{SATsurvey}, and often relies on heuristics for solutions, e.g. the widely used CDCL algorithm~\cite{569607} in modern SAT solvers, that also suffers from exponential run-time in the worst case.

\begin{figure}[t!]
    \centering
    \begin{subfigure}{0.245\textwidth}
         \centering
         \includegraphics[width=0.9\textwidth]{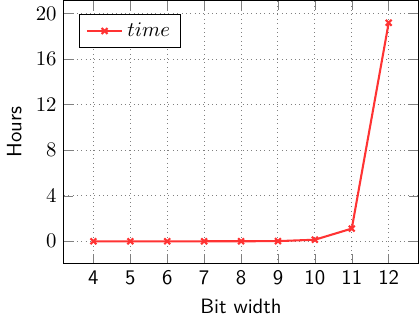}
         \caption{\small Time requirement}
         \label{fig:mul_hours}
    \end{subfigure}%
    \begin{subfigure}{0.245\textwidth}
         \centering
         \includegraphics[width=0.9\textwidth]{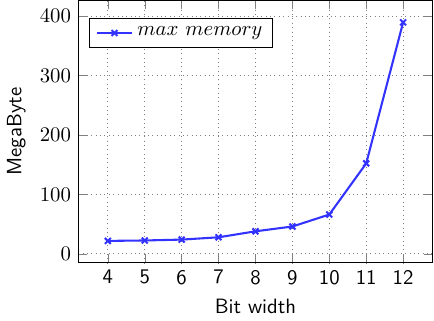}
         \caption{\small Memory requirement}
         \label{fig:mul_meory}
    \end{subfigure}
    \caption{\small Time and memory requirements of SAT based verification of multipliers.}
    \label{fig:multiplier_limitation}
\end{figure}

It is observed that, equivalence checking becomes more easier when the Boolean circuits are of similar structure~\cite{10.1007/978-3-540-24605-3_4}. Major problem is faced when verifying digital circuits with asymmetric structures. Fig.~\ref{fig:multiplier_limitation} shows the run-time and memory requirements needed by Z3 SMT solver for verification of two structurally different multipliers, $M_1: U{SP}+AR+CR$ and $M_2: U{SP}+WT+CL$ for bit-range $4$ to $12$. In both multipliers, the first stage consists of an Unsigned Simple Partial Product Generator ($U{SP}$). They differ in the partial product accumulator and final stage. One uses an Array Multiplier (AR) and Carry Ripple Adder (CR), and the other Wallace Tree multiplier (WT) and Carry Look-ahead Adder (CL). A massive increase in the verification time is observed by  
adding a single bit from 11 to 12.

Quantum computing on the other hand exploiting quantum mechanical phenomena, e.g. superposition and entanglement, seems promising to achieve substantial 
speed-up over some classical algorithms. The emergence of \emph{noisy intermediate-scale quantum} (NISQ) computers~\cite{Cheng2023}  enables one to formulate 
problems taking advantage of such systems.  In this respect, the verification problem is designed as a quantum operator and a quantum version of SAT solver (qSAT) 
is formulated using Grover's search algorithm~\cite{Grover:1996} to find solutions with a quadratic speed-up over classical searching algorithms. 

In this paper we have introduced an ESOP based method for clause generation and corresponding quantum miter circuit interpretation that incurs reduced gate and 
depth overhead. We have also shown that our approach is linear in terms of number of gates present in the circuits considered for verification. Finally, we present 
the qSAT architecture for finding solution to the constructed miter circuits. Further, usage of different reference circuits for verification and corresponding 
resource requirements are also analyzed and presented in the form of case studies. Experiments are conducted to exhibit the advantage of the proposed qSAT approach 
for hardware verification using open-source Qiskit~\cite{Qiskit} platform and IBM quantum computer. The main contributions of the present work can be summarized as follows:
\begin{itemize}
    \item Developing the comprehensive architecture of the qSAT solver by leveraging Grover’s search algorithm. 
\item Optimizing the search space by utilizing ESOP-based clause generation and miter circuit implementation.

\item Assessing the qSAT solver based on gate overhead, solution probability, and operational fidelity across selected benchmarks with specific faults.  
\end{itemize}

The remainder of the paper is structured as follows: 
Section II presents the background on quantum circuit models and the architecture of existing SAT solvers. 
Section III details the proposed ESOP-based clause generation, miter formation, and the qSAT solver architecture, 
including a case study on multiplier and 1-bit full-adder operations. In Section IV, we demonstrate the efficiency of the proposed qSAT solver in terms 
of resource overhead, solution accuracy probability, and operational reliability. Finally, concluding remarks are provided in Section V.

\section{Background}
\subsection{Quantum Circuit}
A quantum circuit consists of multiple qubits, on which a series of gate operations is applied~\cite{Niel:2000}. A circuit consisting of $n$ qubits may contain one or more quantum gates that can operate on $m$ qubits where $m \leqslant n$. The gate operations are represented by unitary square matrices and  qubit states are described as unit vectors representing basis states $|0 \rangle$ or $|1 \rangle$ or as linear sum, $\dirac{\psi} = \alpha |0 \rangle + \beta |1 \rangle$, where $\alpha$ and $\beta$ are complex coefficients and $|\alpha|^2 + |\beta|^2 = 1$. Upon measurement, the state $\dirac{\psi}$ may collapse to one of the basis states, $|0 \rangle$ or $|1 \rangle$, with probabilities $|\alpha|^2$ and $|\beta|^2$ respectively.

\begin{figure}[t!]
    \centering
     \input{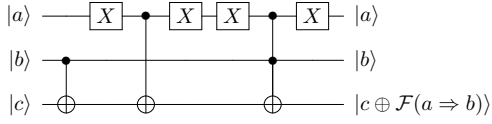}
    \caption{\small Quantum circuit interpretation of a Boolean function,  $\mathcal{F}(a,b) = a \Rightarrow b$.}
    \label{Quantumcircuit}
\end{figure}

Quantum circuits are often described using a set of specific gates. Fig.~\ref{Quantumcircuit} shows a 3-qubit quantum circuit implementing the Boolean function,  $\mathcal{F}:a \Rightarrow b$ using NOT ($X$), CNOT ($CX$) and Toffoli ($C^2X$) gates. Multiple control Toffoli gates, i.e. $C^nX$, for $n\geqslant3$ may also use for quantum circuit realization with their control inputs of positive and negative polarity that are denoted by $\bullet$ and $\circ$, respectively, For execution on a real device like IBM quantum computer, the realized quantum circuit needs to be redescribed using single- and two-qubit primitive gates, e.g. Hadamard ($H$), Phase ($P$), $X$ and $CX$ gates supported by the device.

\subsection{SAT Solvers}
For Boolean Satisfiability (SAT) problem, the solution is to determine an interpretation of the variables, 
also termed as literals, that satisfy the described problem (SAT), if any such interpretation exists; 
otherwise, the problem becomes unsatisfiable (UNSAT)\cite{paraSAToverview}. \emph{Conjunctive normal form} 
(CNF) is used for representation of problem instances, which is a conjunction of clauses where the clauses are defined as the disjunction of literals. For example, $(a \lor b) \land (a \lor c)$ represents a SAT instance with two clauses, $(a \lor b)$ and $(a \lor c)$ where $a=1$, and $b=c=1$ are satisfying literal assignments.      

\begin{figure}[t!]
    \centering
     \includegraphics[scale=0.61]{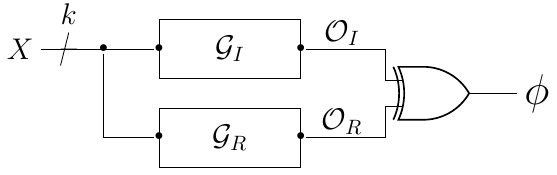}
    \caption{\small Miter circuit for hardware verification of Boolean circuits.}
    \label{miter}
\end{figure}

For hardware verification, SAT-based equivalence checking is performed between a circuit-under-assessment ($\mathcal{G}_I$) and a reference model ($\mathcal{G}_R$). First, a \textit{miter} circuit is assembled using both the circuits as shown in Fig.~\ref{miter} to simplify the equivalence checking into a SAT problem and is then solved by using a SAT solver, e.g. Z3~\cite{de2008z3}. If the circuits are not equivalent, the solver returns SAT and a set of \emph{counter-examples} (CEXs) that consists of the value of literals that satisfy the problem. The solver returns UNSAT when the circuits are functionally equivalent. 

\section{Proposed Quantum SAT Solver}
 
\subsection{ESOP-Based Clause Generation}
In SAT-based verification of synthesized classical circuits, individual gates are often translated into clauses applying schemes like the one presented by Tsytin~\cite{Tseitin1983}. Such translations also introduce auxiliary variables that together with the product clause terms are expensive
for obtaining equivalent unitary realizations. For example, an auxiliary variable $p$ representing 2-input Boolean \emph{OR} operation of the form $q\lor r$ on variables $q$ and $r$, i.e. $\mathcal{F}:p\Leftrightarrow (q\lor r)$ leads to the following \emph{conjunctive normal form} (CNF) clauses:
\begin{align}\label{eq}
p\Leftrightarrow (q\lor r)&=(p\lor \neg q)\land( p\lor \neg r)\land(\neg p\lor q \lor r).    
\end{align}
While \emph{disjunctions} like $\bigvee_{i=1}^kx_i$ has optimal realization in \emph{exclusive sum-of-products} (ESOP) form as $1\oplus \bigwedge_{i=1}^k(\neg x_i)$ for  $k\geqslant 2$ due to~\cite{doi:10.1098/rsta.2019.0161}, a direct interpretation of the CNF corresponds to~\eqref{eq} in the quantum circuit model will still incur large resource 
overheads (i.e., number of qubits, gates and circuit depth) as presented below:

\begin{align}
    \raisebox{1.5cm}{\scalebox{0.8}{
\Qcircuit @C=1.0em @R=0.6em @!R {\\
\lstick{\dirac{p}}    &  \qw                &  \ctrlo{3}                     &  \qw                &  \ctrlo{4}                     &  \qw                &  \ctrl{1}                 &  \qw          &  \qw              &  \rstick{\dirac{p}}\qw \\
\lstick{\dirac{q}}    &  \qw                &  \control\qw                   &  \qw                &  \qw                           &  \qw                &  \ctrlo{1}                    &  \qw          &  \qw          &  \rstick{\dirac{q}}\qw \\
\lstick{\dirac{r}}    &  \qw                &  \qw                           &  \qw                &  \control\qw                   &  \qw                &  \ctrlo{3}                    &  \qw          &  \qw          &  \rstick{\dirac{r}}\qw \\
\lstick{\dirac{0}}    &  \gate{\mathrm{X}}  &  \targ                         &  \qw                &  \qw                           &  \qw                &  \qw                          &  \qw          &  \ctrl{1}     &  \rstick{\dirac{p\lor\neg q}}\qw \\
\lstick{\dirac{0}}    &  \qw                &  \qw                           & \gate{\mathrm{X}}   &  \targ                         &  \qw                &  \qw                          &  \qw          &  \ctrl{1}     &  \rstick{\dirac{p\lor\neg r}}\qw \\
\lstick{\dirac{0}}    &  \qw                &  \qw                           &  \qw                &  \qw                           &  \gate{\mathrm{X}}  &  \targ                        &  \qw          &  \ctrl{1}     &  \rstick{\dirac{\neg p\lor q\lor r}}\qw \\
\lstick{\dirac{0}}  &  \qw                &  \qw                           &  \qw                &  \qw                           &  \qw                &  \qw                          &  \qw          &  \targ        &  \rstick{\dirac{\mathcal{F}: p\Leftrightarrow (q\lor r)}}\qw  
\\
\\
}}} &&   
\end{align}

\begin{figure*}[t!]
    \centering
     \begin{subfigure}[b]{0.18\textwidth}
         \centering
         \includegraphics[scale=0.81]{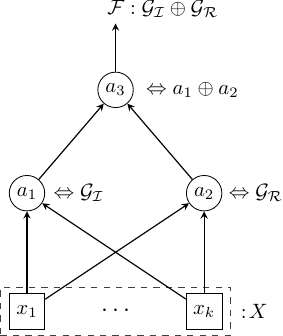}
         \caption{} \label{sing-mit}
     \end{subfigure}\qquad 
     \begin{subfigure}[b]{0.3\textwidth}
         \centering
         \input{figures/single_miter_I}
         \caption{}\label{sing-mitI}
     \end{subfigure}\quad 
     \begin{subfigure}[b]{0.4\textwidth}
         \centering
         \input{figures/single_miter_II}
         \caption{}\label{sing-mitII}
     \end{subfigure}
    \caption{\small An illustration of a quantum miter circuit realization; (a) classical miter circuit $\mathcal{F}$ for doing equivalence checking of Boolean circuits $\mathcal{G_I}$ and $\mathcal{G_R}$; (b) quantum circuit interpretation of $\mathcal{F}$ additionally realizing operations inferred on auxiliary variables $a_1$, $a_2$ and $a_3$; (c) uncomputing the inferences on $a_1$, $a_2$ and $a_3$ applying respective operations in reverse order. }\label{single-miter}
\end{figure*}

Logic expressions like $\mathcal{F}:p \Leftrightarrow \mathcal{G}$ can have different ESOP forms: 
\begin{align} \label{eq1}
    p \Leftrightarrow \mathcal{G} & = \begin{cases} 
                                    (\neg p\land \neg \mathcal{G}) \oplus (p \land \mathcal{G})\text{,} &  \\ 
                                   1 \oplus p \oplus \mathcal{G}\text{.} & 
                                   \end{cases}
\end{align}
This allows one to redefine expression of this form: $p\Leftrightarrow (q\lor r)$ as either $\big(\neg p \land \neg (q \lor r)\big)\oplus \big(p \land ( q \lor r)\big)$ or 
as $p \oplus (\neg q \land \neg r)$, instead of CNF representation~\eqref{eq}, with the corresponding quantum circuit interpretations as follows:

\begin{align}
\raisebox{0cm}{\begin{tabular}{ccc}
\scalebox{0.8}{
\Qcircuit @C=0.5em @R=1.0em @!R {
                        \dstick{\;\;\;\;\;\;\;\;\;\;\;\;\;\;\;\; p \Leftrightarrow (q\lor r)} \\
  \lstick{\dirac{p}}            &  \ctrl{1}            &  \qw &\\
  \lstick{\dirac{q}}            &  \multigate{1}{\bigvee} &  \qw &\dstick{\;\;\;\;\;\;\Large \equiv}\\
  \lstick{\dirac{r}}            &  \ghost{\bigvee}        &  \qw &\\
  \lstick{\dirac{y}}  &  \targ \qwx[-1]      &  \qw & \\
  \\
  \\
}}
&
     \scalebox{0.8}{
\Qcircuit @C=0.5em @R=0.6em @!R { 
 \\
   \lstick{}            &  \qw                &  \qw                    &  \qw                &  \ctrlo{3}                     &  \qw                &  \ctrl{3}                     &  \qw &\\
  \lstick{}            &  \qw                &  \ctrlo{1}               &  \qw                &  \qw                           &  \qw                &  \qw                          &  \qw  &\dstick{\;\;\;\;\;\;\Large \equiv}\\
  \lstick{}            &  \qw                &  \ctrlo{2}               &  \qw                &  \qw                           &  \qw                &  \qw                          &  \qw  &\\
 \lstick{}   &  \qw                &  \qw                     &  \qw                &  \targ                         &  \qw                &  \targ                        &  \qw &\\
  \lstick{\dirac{0}}    &  \gate{\mathrm{X}}  &  \targ                   &  \qw                &  \ctrlo{-1}                    &  \qw                &  \ctrl{-1}                    &  \rstick{\dirac{q\lor r}}\qw \\
 \\
}}
&
     \scalebox{0.8}{
\Qcircuit @C=0.5em @R=1.05em @!R { 
 \\
              &  \ctrl{3}                     &  \qw                &  \qw                     &  \rstick{\dirac{p}} \qw \\
              &  \qw                          &  \qw                &  \ctrlo{1}               &  \rstick{\dirac{q}} \qw \\
              &  \qw                          &  \qw                &  \ctrlo{1}               &  \rstick{\dirac{r}} \qw \\
              &  \targ                        &  \qw                &  \targ                   &  \rstick{\dirac{y\oplus p \Leftrightarrow (q\lor r)}} \qw \\
  \\
}}
\end{tabular}}    &&&&
\end{align}

Thus, considering $\mathcal{G}$ as a $k$-input function of types: OR ($\bigvee_{i=1}^kx_i$), AND ($\bigwedge_{i=1}^kx_i$), and XOR ($\bigoplus_{i=1}^kx_i$) lead to the following ESOP forms:
\begin{align}
    p \Leftrightarrow \bigvee_{i=1}^kx_i & = p \oplus \bigwedge_{i=1}^k\neg x_i\text{,} \\
    p \Leftrightarrow \bigwedge_{i=1}^kx_i & = 1 \oplus p \oplus \bigwedge_{i=1}^kx_i\text{,} \\
    p \Leftrightarrow \bigoplus_{i=1}^kx_i & = 1 \oplus p \oplus \bigoplus_{i=1}^kx_i\text{.} 
\end{align}

Similarly, for the negation ($\neg$) of $\mathcal{G}$, corresponding ESOP forms can be obtained from~\eqref{eq1} by replacing $\mathcal{G}$ with
either $\neg\mathcal{G}$ or $1\oplus \mathcal{G}$, respectively, e.g. an ESOP form for the NOR ($\neg\bigvee_{i=1}^kx_i$) operation will be then:

\begin{align}
    p \Leftrightarrow \neg\Big(\bigvee_{i=1}^kx_i\Big)   & = 1 \oplus p \oplus \bigwedge_{i=1}^k\neg x_i\text{.}
\end{align}

For an arbitrary $k$-input single-target function, $\mathcal{G}(x_1, \dots, x_k)$ an optimal ESOP form can be obtained employing techniques like~\cite{Riener2020} and 
then using~\eqref{eq1} 
the corresponding ESOP forms for the auxiliary variable $p$ realizing $\mathcal{G}$ or $\neg\mathcal{G}$ can be derived. Thus, quantum circuit interpretation of these ESOP-based SAT clauses for each auxiliary variable requires no more than one additional qubit and the implementation can be claimed optimal provided inexpensive ESOP form of $\mathcal{G}$ is employed.    

\subsection{Quantum Miter Circuit Realization}
In order to verify an implemented Boolean circuit $\mathcal{G}_I$, against its corresponding golden reference, $\mathcal{G}_R$, SAT clauses are deduced by introducing auxiliary variables $a_1$ and $a_2$ representing respective functions in the form of $a_1\Leftrightarrow \mathcal{G}_I$ and $a_2\Leftrightarrow \mathcal{G}_R$. Then operational disparity between $\mathcal{G}_I$ and $\mathcal{G}_R$ are asserted by inferring a XOR operation on variables $a_1$ and $a_2$, using another auxiliary variable $a_3$, i.e. $a_3\Leftrightarrow a_1 \oplus a_2$. Fig.~\ref{single-miter} shows the quantum circuit realizations of such \emph{miter} operation for evaluating $\mathcal{F}:\mathcal{G_I}\oplus\mathcal{G_R}$.  

The version of quantum miter implementation additionally providing outcome of operations inferred on auxiliary variables, i.e. $V_\mathcal{F}$ (see Fig.~\ref{sing-mitI}) is economical in terms of number of gates required. This also seems to be useful for asserting the outcome, $\mathcal{F}:\mathcal{G_I}\oplus\mathcal{G_R}$ on answer qubit $\dirac{y}$ 
in the following way: 
\begin{align}
    V_\mathcal{F}\dirac{X,A}\dirac{E}\dirac{y}&= \dirac{X,A}\dirac{{E}'}\dirac{y\oplus\mathcal{G_I}\oplus\mathcal{G_R}}
\end{align}
where $\dirac{X}=\dirac{x_1\dots x_k}$ and $\dirac{A}=\dirac{a_1\dots a_l}$ represent the states of input and auxiliary variables, $\dirac{E}=\dirac{0\dots0}$
indicates the initial state of the ancilla used in computing the inferred operations of the form $a_i\Leftrightarrow \mathcal{G}$, and $\dirac{{E}'}$ denotes the corresponding final ancilla state.

The alternative quantum miter circuit that uncomputes these inferred operations, $a_i\Leftrightarrow \mathcal{G}$ on auxiliary variables, $a_i$, i.e. $U_\mathcal{F}$ (see Fig.~\ref{sing-mitII}) is essential in situations where it becomes necessary to associate only the miter outcome, $\mathcal{F}:\mathcal{G_I}\oplus\mathcal{G_R}$ with the state of input and auxiliary variables as follows: 
\begin{align}
    U_\mathcal{F}\dirac{X,A}&= e^{i\pi(\mathcal{G_I}\oplus\mathcal{G_R})}\dirac{X,A}
\end{align}
where $\dirac{X}=\dirac{x_1\dots x_k}$  and $\dirac{A}=\dirac{a_1\dots a_l}$ denotes the state of input and auxiliary variables. 

\begin{figure}[t!]
    \centering
     \begin{subfigure}[b]{0.18\textwidth}
         \centering
       \includegraphics[width=3.5cm]{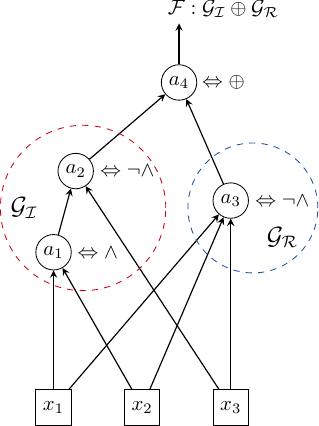}
         \caption{} \label{sing-mit-ex}
     \end{subfigure}\quad 
     \begin{subfigure}[b]{0.26\textwidth}
         \centering
        \input{figures/single_miter_ex}
         \caption{}\label{sing-mit-ex-I}
     \end{subfigure}
    \caption{\small (a) A miter network for verifying the implementation of a $3$-input NAND operation using a pair of $2$-input AND and NAND operations; (b) quantum circuit interpretation of the miter network.}\label{single-miter-ex}
\end{figure}

While both the versions of quantum miter circuit, $V_\mathcal{F}$ and  $U_\mathcal{F}$, differ in terms of required gate operations, the number of required qubits remains identical for both and can be estimated in the following way:
\begin{align}\label{eql}
    \#qubits = |X|+ 2|A|+1
\end{align}
where $|X|$ and $|A|$ denote the number of input and auxiliary variables present in the miter network. For example, Fig.~\ref{single-miter-ex} shows the implementation of a $12$-qubit ($=3 + 2\times 4 + 1$) quantum miter circuit, $V_\mathcal{F}$ for the verification of a $3$-input NAND operation realized using a pair of $2$-input AND and NAND operations.

In order to verify a multi-output Boolean circuit, $\mathcal{G}_I^{1\dots m}$  with the corresponding golden reference, $\mathcal{G}_R^{1\dots m}$ where $m$ ($\geqslant2$) denotes the number of outputs, the miter circuit requires an additional OR operation on auxiliary variables 
$a_i$, i.e. $\mathcal{O}\Leftrightarrow \bigvee_{i=1}^m a_i$ that are used to realize the XOR of the form $a_i\Leftrightarrow b_i\oplus c_i$, where $b_i$ and $c_i$ represent the $i$-th equivalent output, i.e. $b_i\Leftrightarrow \mathcal{G}_I^i$ and  $c_i\Leftrightarrow \mathcal{G}_R^i$. A $V_\mathcal{F}$ type quantum circuit interpretation of such miter, $\mathcal{F}:\bigvee_{i=1}^m \mathcal{G}_I^i \oplus \mathcal{G}_R^i$ can be then:
\begin{align}
    \raisebox{3.4cm}{
    \scalebox{0.72}{
\Qcircuit @C=0.4em @R=0.05em @!R {
                     &&  \dstick{k}   & &&\\
  \nghost{}  & \lstick{\dirac{X}}           & {/}\qw & \gate{\mathcal{G}_I^1} & \qw & \push{\;\dots\;\;}  & \qw  & \gate{\mathcal{G}_I^m} & \gate{\mathcal{G}_R^1} & \qw & \push{\;\dots\;\;} & \qw & \gate{\mathcal{G}_R^m} & \qw & \qw & \qw & \qw & \rstick{\dirac{X}}\qw \\
  \nghost{}  & \lstick{\dirac{b_1}}         &  \qw   & \ctrl{-1}              & \qw & \push{\;\dots\;\;}  & \qw  & \qw                    & \qw                    & \qw & \push{\;\dots\;\;} & \qw  & \qw  & \multigate{1}{\bigoplus} & \qw & \qw & \qw & \rstick{\dirac{b_1}}\qw \\
  \nghost{}  & \lstick{\dirac{c_1}}         &  \qw   & \qw                    & \qw & \push{\;\dots\;\;}  & \qw  & \qw                    & \ctrl{-2}              & \qw & \push{\;\dots\;\;} & \qw  & \qw  & \ghost{\bigoplus} & \qw & \qw & \qw & \rstick{\dirac{c_1}}\qw \\
  \nghost{}  & \lstick{\dots\;}             &        &                        &     & \push{\;\dots\;\;} &      &                         &                        &     & \push{\;\dots\;\;} &      &      &                   &   \push{\;\dots\;\;}   &     &     & \rstick{\;\dots}          \\
  \nghost{}  & \lstick{\dirac{b_m}}         &  \qw   & \qw                    & \qw & \push{\;\dots\;\;}  & \qw  & \ctrl{-4}              & \qw                    & \qw & \push{\;\dots\;\;} & \qw  & \qw  & \qw      & \multigate{1}{\bigoplus}& \qw & \qw & \rstick{\dirac{b_m}}\qw \\
  \nghost{}  & \lstick{\dirac{c_m}}         &  \qw   & \qw                    & \qw & \push{\;\dots\;\;}  & \qw  & \qw                    & \qw                    & \qw & \push{\;\dots\;\;} & \qw  & \ctrl{-5} & \qw & \ghost{\bigoplus} & \qw & \qw & \rstick{\dirac{c_m}}\qw  \\
  \nghost{}  & \lstick{\dirac{a_1}}         &  \qw   & \qw                    & \qw & \push{\;\dots\;\;}  & \qw  & \qw                    & \qw                    & \qw & \push{\;\dots\;\;} & \qw  & \qw  & \ctrl{-5} & \qw & \multigate{2}{\bigvee}& \qw& \rstick{\dirac{a_1}}\qw  \\
  \nghost{}  & \lstick{\dots\;}             &        &                        &     & \push{\;\dots\;\;} &      &                         &                        &     & \push{\;\dots\;\;} &      &      &           &     &                       &    & \rstick{\;\dots}\\
  \nghost{}  & \lstick{\dirac{a_m}}         &  \qw   & \qw                    & \qw & \push{\;\dots\;\;}  & \qw  & \qw                    & \qw                    & \qw & \push{\;\dots\;\;} & \qw  & \qw  & \qw & \ctrl{-3} & \ghost{\bigvee}& \qw & \rstick{\dirac{a_m}}\qw \\
  \nghost{}  & \lstick{\dirac{\mathcal{O}}} &  \qw   & \qw                    & \qw & \push{\;\dots\;\;}  & \qw  & \qw                    & \qw                    & \qw & \push{\;\dots\;\;} & \qw  & \qw  & \qw & \qw & \ctrl{-1} &\ctrl{1}& \rstick{\dirac{\mathcal{O}}}\qw \\
  \nghost{}  & \lstick{\dirac{0}}           &  \qw   & \targ\qwx[-9]          & \qw & \push{\;\dots\;\;}  & \qw  & \qw                    & \qw                    & \qw & \push{\;\dots\;\;} & \qw  & \qw  & \qw & \qw & \qw &\ctrl{2}& \rstick{\dirac{b_1\Leftrightarrow \mathcal{G}_1}}\qw \\
  \nghost{}  & \lstick{\dots\;}             &        &                        &     & \push{\;\dots\;\;} &       & {\vdots}\qwx[-7]       & {\vdots}\qwx[-9]       &     & \push{\;\dots\;\;} &     &  {\vdots}\qwx[-6]  & {\vdots}\qwx[-5] & {\vdots}\qwx[-3] & &\blacksquare& \rstick{\;\dots}\\
  \nghost{}  & \lstick{\dirac{0}}           &  \qw   & \qw                    & \qw & \push{\;\dots\;\;}  & \qw  & \qw                    & \qw                    & \qw & \push{\;\dots\;\;} & \qw  & \qw  & \qw & \qw & \targ\qwx[-3]&\ctrl{1}& \rstick{\dirac{\mathcal{O}\Leftrightarrow \bigvee_{i=1}^m a_i}} \qw \\
  \nghost{}  & \lstick{\dirac{y}}           &  \qw   & \qw                    & \qw & \push{\;\dots\;\;}  & \qw  & \qw                    & \qw                    & \qw & \push{\;\dots\;\;} & \qw  & \qw  & \qw & \qw &\qw &\targ & \rstick{\dirac{y\oplus \mathcal{F}}}\qw \\
  \\
}}
} &&&&&&    
\end{align}

A miter $\mathcal{F}$ may have more than one \emph{counter-example} (CEX) of the following form: 
\begin{align}\label{eq2}
    \textit{CEX}=\Big(\bigwedge_{i=1}^kx_i=v_i\Big)\land\Big(\bigwedge_{j=1}^la_j=v_j\Big)
\end{align} 
where $x_i\in X$ and $a_j\in A$ indicate input and auxiliary variables respectively, and $v_i, v_j\in\{0,1\}$ are their respective binary values. In order to exclude $m$ such CEX from the miter outcome, $\mathcal{F}$ one can consider the following augmented ESOP form:
\begin{align}\label{eqa}
    \widehat{\mathcal{F}} = \mathcal{F} \oplus \bigoplus_{i=1}^m\textit{CEX}_{i}
\end{align}
where $\textit{CEX}_{i}$ indicates the $i$-th counter-example.
The quantum circuit interpretation of such augmented miter, $\widehat{\mathcal{F}}$ can be obtained by realizing the miter, $\mathcal{F}$  in either of $V_\mathcal{F}$ or $U_\mathcal{F}$ form 
followed by placing multiple-controlled-$X$ operation, or $C^{k+l}X$ in short, with the qubits representing input ($X$) and auxiliary ($A$) variables as the control 
and the same answer qubit, $y$ that is used for realizing outcome of $\mathcal{F}$ (see Fig.~\ref{single-miter}) as the target in the following way: 
\begin{align}
    C^{k+l}X\dirac{X,A}\dirac{y} =\dirac{X,A}\dirac{y\oplus X \land A}
\end{align}
where the polarity of control qubits, $x_i\in X$ or $a_i\in A$  is similar to the logic states, $v_i\in \{0,1\}$ of the corresponding satisfying assignment, and for optimal realization of such $C^nX$ gate some of the reported compilation techniques ( e.g.,~\cite{PhysRevA.93.022311,Amy:2013,Baren:1995}) can be exploited. 
A Grover's search based approach is outlined next to find one of these CEXs as miter solution. 

\begin{figure}[t!]
    \centering
    \includegraphics[width=7.5cm]{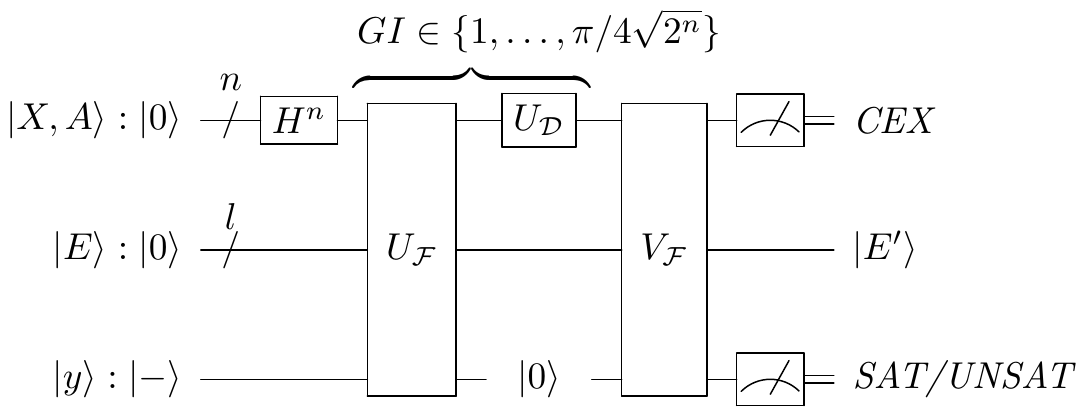}
    \caption{\small Outline of the qSAT Algorithm.}\label{qsat-arc}
\end{figure}

\subsection{qSAT Architecture}

In order to find a \emph{counter-example} (CEX) for the constructed quantum miter circuit, initially Grover's search algorithm~\cite{Grover:1996} is exploited. 
The $U_\mathcal{F}$ version of the quantum miter circuit is used as the oracle to conduct the \emph{phase inversion} operation. This is followed by the application of \emph{Grover's diffusion} ($U_\mathcal{D}$) operator to amplify the probability of all possible CEXs.
After placing the qubits representing the input and auxiliary variables denoted by $X$ and $A$, respectively in superposition, i.e. $H^{\otimes n}\dirac{X,A}$ where: 
\begin{align}\label{eqn}
n=|X|+|A|\text{,}    
\end{align}  
the oracle,  $U_\mathcal{F}$ and the diffuser, $U_\mathcal{D}$ are applied in sequence for a maximum of $O(\sqrt{2^n})$ times. 
With the amplitude of CEXs are being amplified, the quantum miter circuit is executed one more time before applying quantum measurement operation to ensure $\textit{SAT/UNSAT}$ miter outcome.  For this final miter evaluation, $V_\mathcal{F}$ version can be used for optimal realization without affecting the desired outcome. Fig.~\ref{qsat-arc} shows the outline of the qSAT algorithm where label $E$ indicates the additional qubits that are used as ancilla such that $|E|=|A|=l$ due to~\eqref{eql} and their final states denoted by $\dirac{E'}$ are ignored from the measured outcome. In a noise-free or fault-tolerant quantum processing environment, the additional miter ($V_\mathcal{F}$) execution in the qSAT algorithm eliminates the need of classical verification of the measured CEX.

A prior knowledge about the number of CEXs, (say, $m$) can reduce the number of executions of \emph{Grover's iteration} (GI), i.e. $U_\mathcal{F}$ and $U_\mathcal{D}$ in qSAT algorithm by setting $GI={\pi}/{4}\sqrt{2^n/m}$ (see Fig.~\ref{qsat-arc}). While $m$ remains unknown in most of the cases and can only be obtained by employing techniques like quantum counting~\cite{10.1007/BFb0055105} in order to achieve the quadratic speed up over classical exponential runtime, $O(2^n)$.  In present scenario we consider ${\pi}/{4}\sqrt{{2^n}/{m}} + 1$ executions of the miter for obtaining a verified CEX.   

In order to exclude one or more CEXs from the evaluation outcome, the miter, $U_\mathcal{F}$ used in the qSAT algorithm 
(see Fig.~\ref{qsat-arc}) must be replaced by the corresponding augmented miter, $U_{\widehat{\mathcal{F}}}$ constructed 
based on Eqn.~\eqref{eqa}. This reduces the number of remaining CEXs, $m'$ ($< m$) which results in increased Grover's iteration.  
The miter, $V_\mathcal{F}$  used for final verification of CEXs should be kept unaltered. 

Finally, the number of Grover's iteration ${\pi}/{4}\sqrt{{2^n}/{m}}$ increases rapidly with $n$ ($\geqslant 10$) that according to~\eqref{eqn}, can be further reduced by minimizing the number of considered auxiliary ($A$) variables while keeping the number of input ($X$) variables unchanged. A possible solution towards this in the form of a case study is presented next.   

\begin{figure}[t!]
    \centering
     \begin{subfigure}[b]{0.18\textwidth}
         \centering
       \includegraphics[width=3.5cm]{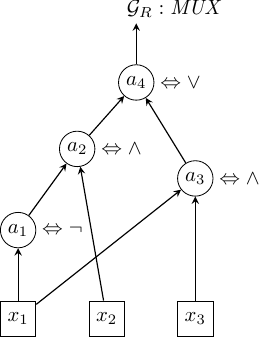}
         \caption{} \label{muxI}
     \end{subfigure}\quad 
     \begin{subfigure}[b]{0.18\textwidth}
         \centering
        \includegraphics[width=2.85cm]{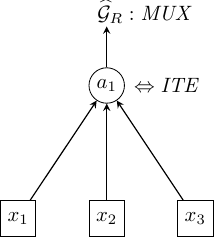}
         \caption{}\label{muxII}
     \end{subfigure}
     
    \caption{\small (a) A reference multiplexer clause network realization using $4$ auxiliary variables; (b) corresponding realization based on single auxiliary variable.}\label{ref-mux}
\end{figure}

\subsection{qSAT Optimization: Case Study}
For multiplexer and $1$-bit full-adder realizations of different reference model, $\mathcal{G}_R$ are analyzed in terms of the number of required Grover's iterations besides qubits, gates, and depths.
\paragraph*{Multiplexer}
The following description of a $2$-input multiplexer:
\begin{align} \label{eqif}
    \textit{MUX}(x_1, x_2, x_3) &= (\neg x_1\land x_2) \lor (x_1\land x_3)
\end{align} 
can have the ESOP form:
\begin{align}\label{eqif2}
     \textit{ITE}(x_1, x_2, x_3)&= x_2 \oplus (x_1 \land x_2) \oplus (x_1 \land x_3)
\end{align}
where the logic states of $x_1$ decides either of the logic sate of $x_2$ and $x_3$ as final output, i.e.  \emph{If} $x_1$ \emph{is high Then output} $x_3$ \emph{Else} $x_2$. Fig.~\ref{ref-mux} shows the corresponding multiplexer clause networks based on~\eqref{eqif} and~\eqref{eqif2}. The usage of~\eqref{eqif} in ESOP based clause generation for $\mathcal{G}_R$ requires $4$ auxiliary variables and respective quantum circuit interpretation requires additional $8$ ($=2\times 4$, due to~\eqref{eql}) qubits, whereas using a single auxiliary variable the same, denoted by $\widehat{\mathcal{G}}_R$ (see Fig.~\ref{muxII}) can be achieved using only $2$ additional qubits when~\eqref{eqif2} is exploited. 
This besides providing inexpensive miter ($U_\mathbf{F}/V_\mathbf{F}$) and diffuser ($U_\mathcal{D}$) realizations in terms of qubits, gates and depths, 
the number of Grover's iterations get reduced by a factor of $\sqrt{2^{-3}}$ when $\widehat{\mathcal{G}}_R$ is used as reference model in miter construction.

\begin{figure}[t!]
    \centering
     \begin{subfigure}[b]{0.22\textwidth}
         \centering
       \includegraphics[width=4.0cm]{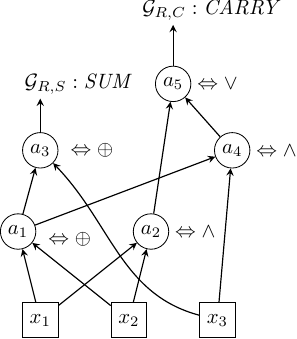}
         \caption{} \label{fa}
     \end{subfigure}\quad 
     \begin{subfigure}[b]{0.20\textwidth}
         \centering
        \includegraphics[width=3.9cm]{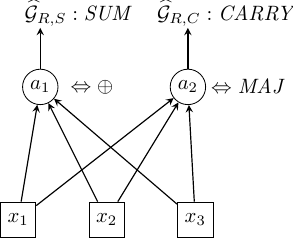}
         \caption{}\label{faII}
     \end{subfigure}
     
    \caption{\small (a) A reference $1$-bit full-adder clause network with $5$ auxiliary variables; (b) corresponding $2$ auxiliary variables based realization.}\label{ref-mux}
\end{figure}

\paragraph*{1-Bit Full-Adder}
The functionality of a full-adder is often represented in the following way:
\begin{align}
    \textit{SUM} &=  (x_1 \oplus x_2) \oplus x_3 \label{fas}\text{ and}\\
    \textit{CARRY} &= ((x_1 \oplus x_2) \land x_3 ) \lor (x_1 \land x_2)\label{fac}\text{,}
\end{align}
where $x_1$, $x_2$ and $x_3$ are the inputs used in obtaining the two outputs, $\textit{SUM}$ and $\textit{CARRY}$. Clause realization for $\mathcal{G}_R$ based on~\eqref{fas} and~\eqref{fac} requires $5$ auxiliary variables as shown in Fig.~\ref{fa} and corresponding quantum circuit interpretation demands $10$ additional qubits due to~\eqref{eql}. Considering the $\textit{CARRY}$ as majority ($\textit{MAJ}$) operation and $\textit{SUM}$ as 3-input XOR operation, we can have the following ESOP form of the full-adder:
\begin{align}
    \textit{SUM} &=  x_1 \oplus x_2 \oplus x_3 \label{fas2}\text{ and}\\
    \textit{MAJ} &= (x_1 \land x_2) \oplus (x_1 \land x_3) \oplus (x_2 \land x_3) \label{fac2}\text{,}
\end{align}
that requires only $2$ auxiliary variables as shown in Fig.~\ref{faII} and leads to quantum circuit realization using $4$ additional qubits. Thus for quantum miter realization the use of reference model ($\widehat{\mathcal{G}}_R$) based on~\eqref{fas2} and~\eqref{fac2} 
also bring downs the number of Grover's iterations by a factor of $\sqrt{2^{-3}}$. 

\section{Experimental Evaluation}

For assessing the effectiveness of the proposed qSAT based verification approach, the open-source Qiskit~\cite{Qiskit} platform  
together with IBM quantum computer are used for construction of quantum miter circuits, evaluation of quantum resource requirements of qSAT solver and analyzing clause executions.

{ \setlength{\tabcolsep}{4pt}
\begin{table}[htbp]
  \caption{\small Boolean functions considered for qSAT experiments. }
  \label{tab1}
  \begin{tabular}{llcc} \toprule 
  Name                  & Description & Faults & $\#CEX$\\\midrule
  AND                   & $(x_1\;{\color{red}\land}\;x_2)\; \land x_3$ & $\neg\lor$ & $2$ \\
  NAND                  & $(x_1\;{\color{red}\land}\;x_2)\; \neg\!\land x_3$ & $\lor$ & $2$ \\
  OR                    & $(x_1\;{\color{red}\lor}\;x_2)\;  \lor x_3$ & $\oplus$ & $1$ \\
  NOR                   & $(x_1\;{\color{red}\lor}\;x_2)\; \neg\!\lor x_3$ & $\neg\oplus$ & $3$ \\
  XOR                   & $(x_1\;{\color{red}\oplus}\;x_2)\; \oplus x_3$ & $\neg\oplus$ & $8$ \\
  XNOR                  & $(x_1\;{\color{red}\oplus}\;x_2)\; \neg\!\oplus x_3$ & $\land$ & $6$ \\
  MUX                   & $(\neg x_1\land x_2)\; \lor \;(x_1\;{\color{red}\land}\;x_3)$ & $\neg\land$ & $6$ \\
  CARRY                 & $\big((x_1\oplus x_2)\; \land \; x_3\big)\; \lor \;(x_1\;{\color{red}\land}\;x_2)$ & $\neg\lor$ & $4$ \\
  \multirow{ 2}{*}{FA}  & $(x_1\;{\color{red}\oplus^1}\;x_2)\; \oplus x_3$, & \multirow{2}{*}{$\neg\oplus^1$, $\neg\lor^2$ } & \multirow{ 2}{*}{$8$} \\
                        & $\big((x_1\;{\color{red}\oplus^1}\;x_2)\; \land \; x_3\big)\; \lor \;(x_1\;{\color{red}\land^2}\;x_2)$ & & \\ \bottomrule
  
  \end{tabular}
  
  \end{table}
}

\subsection{Considered Boolean Functions}
 For equivalence checking, Boolean functions with the following realizations using: (i) inverter ($\neg$) and $2$-input logic gates, and (ii) $3$-input logic gates including multiplexer ($\textit{ITE}$)  and carry ($\textit{MAJ}$) operations are considered as reference models and are denoted by $\mathcal{G}_R$ and $\widehat{\mathcal{G}}_R$, respectively. The corresponding faulty versions, i.e. $\mathcal{G}_F$, are then obtained by replacing one or more $2$-input gates from the $\mathcal{G}_R$ type reference networks. Table~\ref{tab1} shows the list of Boolean functions considered for the experiments, along with their logic descriptions in terms of $2$-input gates ($\mathcal{G}_R$), inserted faults replacing the logic operation highlighted in red color to obtain faulty realization ($\mathcal{G}_F$) and number of \emph{counter-example} (CEX) that results in evaluating the corresponding miter of the form $\mathcal{G}_F\oplus \mathcal{G}_R$ using Z3 SMT solver~\cite{de2008z3}. Based on two different reference models $\mathcal{G}_R$ and $\widehat{\mathcal{G}}_R$, the ESOP based clauses are generated for the respective miters: $\mathcal{G}_F\oplus \mathcal{G}_R$ and $\mathcal{G}_F\oplus \widehat{\mathcal{G}}_R$.      
 
{ \setlength{\tabcolsep}{6pt}
\begin{table*}[t!]
  \caption{\small Analysis of quantum resources for qSAT solving of quantum miters constructed using $\mathcal{G}_R$ and $\widehat{\mathcal{G}}_R$ type reference models.}
  \label{tab2}
  \centering
  \begin{tabular}{l|rrrrrr|rrrrrr|rrrr} \toprule 
 & \multicolumn{6}{|c|}{$\textit{qSAT}{(\mathcal{G}_F\oplus \mathcal{G}_R)}$} & \multicolumn{6}{c|}{$\textit{qSAT}{(\mathcal{G}_F\oplus \widehat{\mathcal{G}}_R)}$} & \multicolumn{4}{c}{Improv.($\%$)} \\ \cmidrule(lr){2-7}\cmidrule(lr){8-13}\cmidrule(lr){14-17} 
Name & \multicolumn{1}{c}{$\#q$} & \multicolumn{1}{c}{$|A|$} & \multicolumn{1}{c}{$GI$} & \multicolumn{1}{c}{$\#CX$} & \multicolumn{1}{c}{$\#U$} & \multicolumn{1}{c|}{$\#D$} & \multicolumn{1}{c}{$\#q$} & \multicolumn{1}{c}{$|A|$} & \multicolumn{1}{c}{$GI$} & \multicolumn{1}{c}{$\#CX$} & \multicolumn{1}{c}{$\#U$} & \multicolumn{1}{c|}{$\#D$} & 
$\#q$ & $\#CX$ & $\#U$ & $\#D$ \\ \midrule   
AND & 14 & 5 & 6 & 969 & 1669 & 1542 & 12 & 4 & 4 & 418 & 534 & 787 & 14.29 & 56.86 & 68.00 & 48.96 \\ 
NAND & 14 & 5 & 6 & 969 & 1630 & 1530 & 12 & 4 & 4 & 418 & 531 & 787 & 14.29 & 56.86 & 67.42 & 48.56 \\
OR & 14 & 5 & 8 & 1197 & 2061 & 2063 & 12 & 4 & 6 & 628 & 808 & 1137 & 14.29 & 47.54 & 60.80 & 44.89 \\
NOR & 14 & 5 & 5 & 777 & 1349 & 1330 & 12 & 4 & 3 & 340 & 443 & 618 & 14.29 & 56.24 & 67.16 & 53.53 \\
XOR(SUM) & 14 & 5 & 3 & 413 & 624 & 765 & 12 & 4 & 2 & 239 & 247 & 385 & 14.29 & 42.13 & 60.42 & 49.67 \\
XNOR & 14 & 5 & 3 & 441 & 680 & 845 & 12 & 4 & 2 & 229 & 251 & 387 & 14.29 & 48.07 & 63.09 & 54.20 \\   
MUX & 22 & 9 & 13 & 3433 & 5608 & 5651 & 16 & 6 & 5 & 1011 & 1692 & 1649 & 27.27 & 70.55 & 69.83 & 70.82 \\ 
CARRY & 22 & 9 & 16 & 4245 & 7114 & 7123 & 16 & 6 & 6 & 1284 & 2088 & 2067 & 27.27 & 69.75 & 70.65 & 70.98 \\
FA & 30 & 13 & 45 & 16413 & 26781 & 26741 & 24 & 10 & 16 & 4941 & 7776 & 7742 & 20.00 & 69.90 & 70.96 & 71.05 \\
\bottomrule
    
  \end{tabular}
  
  \end{table*}
}
\subsection{Resources Assessment for qSAT Solving}
For the construction of qSAT network, initially quantum circuit interpretation of the miters, $\mathcal{G}_F\oplus \mathcal{G}_R$ and $\mathcal{G}_F\oplus \widehat{\mathcal{G}}_R$ are obtained. Then based on the number of input variables ($|X|$), auxiliary variables ($|A|$) and counter-examples ($\#CEX$), Grover's iteration ($\textit{GI}$) is derived in the following way:
\begin{align}
    \textit{GI} &= \frac{1}{2}\sqrt{\frac{2^{|X|+|A|}}{\#CEX}}.
\end{align}
The obtained qSAT network is compiled using Qiskit \emph{transpiler} setting $\{CX, X, P, H\}$ as \emph{basis\_gates} and $\textit{optimization\_level}=3$. For optimal realization, all $C^nX$ type gates of size $n+1$ qubits ($n\geqslant 2$) from the qSAT network are compiled either in \emph{v-chain} or \emph{v-chain-dirty} mode depending on the availability of remaining $n-2$ qubits in the network and their current state.   
Table~\ref{tab2} shows an analysis of quantum resources required by the qSAT network for both versions of the quantum miters, $\mathcal{G}_F\oplus \mathcal{G}_R$ and $\mathcal{G}_F\oplus \widehat{\mathcal{G}}_R$. Since for all the Boolean functions the number of inputs remain unchanged, i.e. $|X|=3$, the usage of reference models of the form $\widehat{\mathcal{G}}_R$ results in less number of auxiliary variables and Grover's iterations. This leads to inexpensive qSAT networks in terms of qubits ($\#q$), single- and two-qubit gates ($\#U$ and $\#CX$, respectively), and depth ($\#D$) compared to the qSAT networks obtained considering $\mathcal{G}_R$ as the reference model.

\subsection{Evaluation of qSAT Outcome}
\begin{figure}[t!]
     \centering

     \scalebox{0.8}{
\begin{tikzpicture}
  \begin{axis}[
    ylabel={Probability of SAT},
    xmin=0, xmax=9,
    ymin=0, ymax=1.0,
    xtick={1,2,3,4,5,6,7,8},
    ytick={0,0.25,0.5,0.75,1},
    xticklabels={AND, NAND, OR, NOR, XOR, XNOR, MUX, CARRY},
    xticklabel style={rotate=90},
    legend style={at={(0.97,0.65)}},
    ymajorgrids,
  ]

\addplot[color=teal,mark=*] coordinates {
    (1, 0.830) 
    (2, 0.858)
    (3, 0.761)
    (4, 0.859)
    (5, 0.879)
    (6, 0.758)
    (7, 0.754)
    (8, 0.741)
  };
  \addlegendentry{$M:\mathcal{G}_F\oplus \mathcal{G}_R$}
  \addplot[color=blue,mark=square*] coordinates {
    (1, 0.801)
    (2, 0.829)
    (3, 0.830)
    (4, 0.769)
    (5, 0.914)
    (6, 0.799)
    (7, 0.872)
    (8, 0.843)
  };
  \addlegendentry{$\widehat{M}:\mathcal{G}_F\oplus \widehat{\mathcal{G}}_R$}

  \addplot[color=red,mark=triangle*] coordinates {
    (1, 0.029)
    (2, 0.029)
    (3, 0.069)
    (4, 0.090)
    (5, 0.036)
    (6, 0.035)
    (7, 0.118)
    (8, 0.102)
  };
  \addlegendentry{$|M-\widehat{M}|\quad\;\;$}
  
  \end{axis}
\end{tikzpicture}}          
     \caption{\small  qSAT accuracy in providing SAT for miters, $\mathcal{G}_F\oplus \mathcal{G}_R$ and $\mathcal{G}_F\oplus \widehat{\mathcal{G}}_R$.}
     \label{fig:result} 
\end{figure}
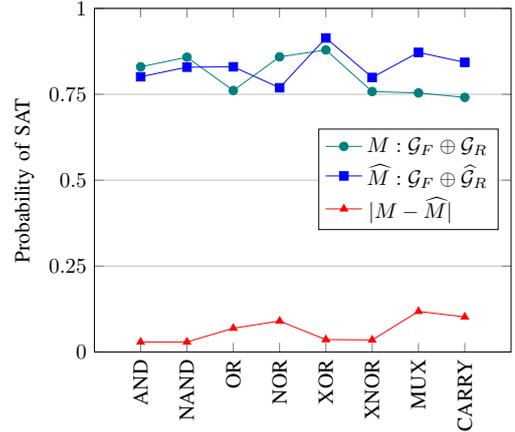

For evaluation of qSAT networks we consider Qiskit \emph{Aer} simulator as an ideal environment. The probability of SAT outcome for each miter, $M\in\{\mathcal{G}_F\oplus \mathcal{G}_R, \mathcal{G}_F\oplus \widehat{\mathcal{G}}_R \}$, is computed in the following way:
\begin{align}\label{eqprob}
    P_M(SAT) &= \frac{1}{\#run}\sum_{i=1}^n P(CEX_i) 
\end{align}
where $\#run$ indicates the number of solver executions for $M$ and $P(CEX_i)$ denotes the probability of obtaining the corresponding $i$-th counter example.
Fig.~\ref{fig:result} shows a comparison of execution results for all the miter circuits of types, $\mathcal{G}_F\oplus \mathcal{G}_R$ and $\mathcal{G}_F\oplus \widehat{\mathcal{G}}_R$. It can be observed that for miter, $\mathcal{G}_F\oplus \widehat{\mathcal{G}}_R$, qSAT solver provides similar results, i.e. $P(SAT)\geqslant 75\%$ utilizing comparatively less quantum resources than the miter, $\mathcal{G}_F\oplus \mathcal{G}_R$.   
Due to experiencing longer simulator run-time, the results for FA are not considered in the present analysis.  

\subsection{Analysis of Reference Model Fidelity}
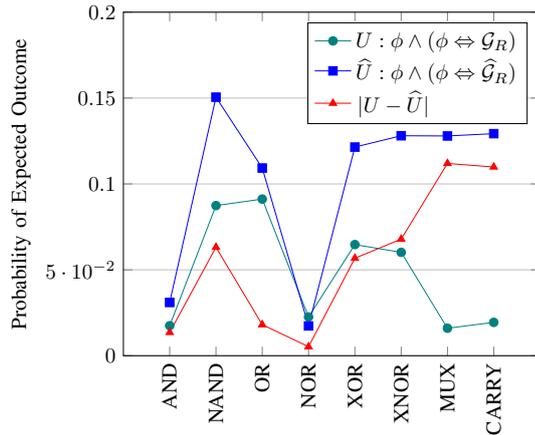
\begin{figure}[htbp!]
     \centering

     \scalebox{0.8}{
\begin{tikzpicture}
  \begin{axis}[
    ylabel={Probability of Expected Outcome},
    xmin=0, xmax=9,
    ymin=0, ymax=0.2,
    xtick={1,2,3,4,5,6,7,8},
    ytick={0,0.05,0.1,0.15,0.2},
    xticklabels={AND, NAND, OR, NOR, XOR, XNOR, MUX, CARRY},
    xticklabel style={rotate=90},
    legend style={at={(0.97,0.97)}},
    ymajorgrids,
  ]

\addplot[color=teal,mark=*] coordinates {
    (1, 0.0174560546875) 
    (2, 0.087402344)
    (3, 0.0911865234375)
    (4, 0.0225830078125)
    (5, 0.064697265625)
    (6, 0.0601806640625)
    (7, 0.015991211)
    (8, 0.01940918)
  };
  \addlegendentry{$U:\phi\land(\phi\Leftrightarrow \mathcal{G}_R)$}
  \addplot[color=blue,mark=square*] coordinates {
    (1, 0.031005859375)
    (2, 0.1505126953125)
    (3, 0.1092529296875)
    (4, 0.017333984375)
    (5, 0.1214599609375)
    (6, 0.1280517578125)
    (7, 0.127929688)
    (8, 0.129272461)
  };
  \addlegendentry{$\widehat{U}:\phi\land(\phi\Leftrightarrow \widehat{\mathcal{G}}_R)$}

  \addplot[color=red,mark=triangle*] coordinates {
    (1, 0.0135498046875)
    (2, 0.063110352)
    (3, 0.01806640625)
    (4, 0.0052490234375)
    (5, 0.0567626953125)
    (6, 0.06787109375)
    (7, 0.111938477)
    (8, 0.109863281)
  };
  \addlegendentry{$|U-\widehat{U}|\qquad\qquad$}
  
  \end{axis}
\end{tikzpicture}}          
     \caption{\small Fidelity of reference models, $\mathcal{G}_R$ and $\widehat{\mathcal{G}}_R$ on a IBM quantum computer.}
     \label{fig:result2} 
\end{figure}

To assess the fidelity of both the versions of reference models, $\mathcal{G}_R$ and $\widehat{\mathcal{G}}_R$, the following CNF forms: 
\begin{align}
    U&=\phi\land(\phi\Leftrightarrow \mathcal{G}_R) \\
    \widehat{U} &= \phi\land(\phi\Leftrightarrow \widehat{\mathcal{G}}_R)
\end{align} 
are considered for quantum circuit interpretation. Initially, putting input ($X$) and auxiliary ($A$) qubits in superposition applying Hadamard ($H$) gates, the CNFs, $U$ and $\widehat{U}$ are executed in isolation on a IBM quantum computer, \emph{ibm\_brisbane} with parameters: $\textit{basis\_gates} = \{ECR, RZ, SX, X\}$ and $\textit{routing\_method}=\textit{`sabre'}$, and their probability of expected outcomes, computed using approach similar to~\eqref{eqprob}, are presented in Fig.~\ref{fig:result2}. The $\widehat{\mathcal{G}}_R$ type reference model shows higher operational fidelity for almost all Boolean functions. 

\section{Conclusion}
In this paper we introduce an improved quantum version of the Boolean Satisfiability solver known as qSAT. Initially the clauses are generated using ESOP based method, then the miter circuit is proposed. The proposed method results in circuit generation with fewer qubits, which is also linear in terms of number of gates present in the circuit. Finally, the qSAT architecture is proposed which exploits the benefit of Grovers search algorithm. Experimental results reveal that our proposed qSAT architecture require less quantum resources in terms of gates and depth. Fidelity analysis is also performed on the reference models. To the best of our knowledge this is a first attempt towards design of a complete qSAT solver.

\bibliographystyle{IEEEtran}
\bibliography{assets/quantum-2022, assets/quantum}

\end{document}